\documentclass{article}

\usepackage{arxiv}
\usepackage[utf8]{inputenc} 
\usepackage[T1]{fontenc}    
\usepackage{hyperref}       
\usepackage{url}            
\usepackage{booktabs}       
\usepackage{amsfonts}       
\usepackage{nicefrac}       
\usepackage{microtype}      
\usepackage{xcolor}
\usepackage{lipsum}
\usepackage{amsmath}
\usepackage{float}
\usepackage{graphicx}
\usepackage{caption}
\usepackage{subcaption}
\usepackage{empheq}
\usepackage{textcomp}
\usepackage{cite}

\title{Analog Optical Implementation of Spatial High Pass Filtering Using Evanescent Waves}
\author{Ali P. Vafa \\
Department of Electrical Engineering\\
Sharif University of Technology\\
Tehran, Iran
\And Parisa Karimi \\
Department of Electrical Engineering\\
Sharif University of Technology\\
Tehran, Iran
\And Amin Khavasi\thanks{khavasi@sharif.edu} \\
Department of Electrical Engineering\\
Sharif University of Technology\\
Tehran, Iran}
\date{June 2019}

\begin{document}

\maketitle

\begin{abstract}
We introduce optical polarization-insensitive high pass filters based on total internal reflection of light at the interface of two dielectric media (1D) and Bragg reflection of a multilayer stack (2D) in transmission mode. The wavevectors in the stopband become coupled to evanescent waves in our design, rather than the zero of a narrow-band resonant mode. This provides remarkable resolution enhancement for edge detection applications. Rigorous analysis based on plane wave expansion is carried out and the results are verified by full-wave numerical simulation. Also for the case of multilayer structure, the thickness of layers is tuned using an optimization algorithm to represent a better approximation of an ideal high pass filter. The application of the designed high pass filters for edge detection of input field profiles is demonstrated for both 1D and 2D operations. The proposed devices are compact and no Fourier lens is required, since the operator is directly implemented in the spatial Fourier domain.
\end{abstract}


\section{Introduction}
Digital computers, comprising universal logic gates, are tremendously versatile. This versatility associated with the increasing integrability of digital electronics due to Moore’s law has led these computers to prevail over their analog counterparts in the last few decades. However, in the case of some specific problems such as modeling of complex and nonlinear systems or processing of large amounts of data, computation time and power are still serious restrictions of digital computing. Optical phenomena are fast enough and offer novel approaches to overcome the mentioned restrictions about system speed and power consumption \cite{Solli2015,Caulfield2010}. Computational metamaterials, a recent field of study on spatial optical analog computing and a successor of lens-based Fourier Optics \cite{goodman2005introduction}, uses the potential of ultra-compact metamaterials and metasurfaces to perform basic mathematical operations on optical wavefronts. The parallel-processing feature of these passive optical devices promises a high-throughput real-time framework of computation with ultra-low power consumption for all-optical image processing applications. Different linear mathematical operations such as 1st / 2nd order differentiation, Laplacian, integration, and convolution have been investigated and several optical devices have been proposed to implement such operations \cite{recentadvances,Silva2014,Pors2015,AbdollahRamezani2015,Chizari2016,Abdollahramezani2017,Bykov2014,Zhu2017,Zangeneh-Nejad2017,Bykov:18,Kwon2018,Guo2018,Guo2018:diff,Saba2018,Bezus2018,Youssefi2016,Zangeneh-Nejad2018,Nesterenko2018,Zhu2019}. A wide range of optical phenomena has been utilized for realization of these operations. The potential of artificially engineered metamaterials to arbitrarily manipulate wavefronts in deeply sub-wavelength scales \cite{Yu2011,Kildishev2013,Monticone2013} together with gradient-index (GRIN) lenses laid the foundation to design optical analog operators in the spatial domain \cite{Silva2014,Pors2015,AbdollahRamezani2015,Chizari2016,Abdollahramezani2017}. On the other hand, the angular-dependant response of much simpler optical structures based on resonant \cite{Bykov2014,Zhu2017,Zangeneh-Nejad2017,Bykov:18,Kwon2018,Guo2018,Guo2018:diff,Saba2018,Bezus2018} or non-resonant \cite{Youssefi2016,Zangeneh-Nejad2018,Nesterenko2018,Zhu2019} optical phenomena enabled practical solutions of performing mathematical operations in the wavevector domain. However, there has been less concentration on spatial high pass filtering as a basic linear transfer function. A 2D high pass filter, proposed in \cite{Guo2018}, is based on guided-mode resonance in a photonic crystal structure. Nevertheless, the fabrication process of this design is relatively complicated due to nano-lithography concerns. Yet, the realized transfer function does not exhibit a sharp transition at the cutoff frequency, as one may expect of an ideal high pass filter.

In this paper, optical structures are proposed to apply spatial high pass filtering (HPF) on input beam profiles with 1D and 2D functionality in the transmission mode which facilitates the integration of these devices with other consequent processing units. The performance of these structures does not depend on the polarization of incident field and can be employed for TE or TM incidence.

For the case of 1D HPF, the proposed structure is based on the total internal reflection (TIR) of light at the interface of two dielectric media where no Fourier lenses are required. According to Fresnel coefficients of reflection and transmission at such an interface, the critical incident angle for TIR to take place does not depend on the wavelength nor the polarization of incident beam. Assuming low-dispersion dielectrics, this makes the desired HPF operation to be polarization-insensitive in a wide range of wavelengths, in contrast to similar operators based on plasmonic resonance \cite{Zhu2017} and Brewster effect \cite{Youssefi2016,Nesterenko2018}. Moreover, the non-resonant nature of total internal reflection provides a remarkable spatial bandwidth for applications such as edge detection and permits resolving edges in the input profile with a finer quality \cite{Canny1986,Marr1980}.

Next, a multilayer dielectric structure is investigated to implement a 2D isotropic high pass filter in the wavevector domain. An isotropic transfer function is one that its amplitude and phase depend only on the magnitude of the transverse wavevector and not its direction. This is a desirable feature in image processing applications and emerges from the transverse symmetry of the multilayer structure. It is shown that by adjusting the thickness of layers through an optimization algorithm, the optical transfer function of the structure exhibits a more acceptable HPF functionality. Finally, the application of the designed high pass filter for detecting sharp edges in the incident wavefront along both transverse axes is illustrated by numerical simulations.

These notably simple and compact optical devices can be utilized for analog optical image processing applications where real-time processing of large amounts of data  with low power consumption is desired.

\section{Theory}
A general 3D beam profile propagating along the $+z$ axis in a medium having refractive index $n$ is thoroughly recognized by its vector-valued electric field distribution $\mathbf{E}(\mathbf{r})$, where $\mathbf{r}=(x,y,z)$ is the position vector in space. This field distribution can simply be reformulated as a superposition of $TE_z$ and $TM_z$ plane waves using Fourier analysis, as follows \cite{Bykov2014}:
\begin{equation} \label{equ:FourierDecomp}
    \begin{split}
    \mathbf{E}(x,y,z) &= \frac{1}{(2\pi)^2}\int\int \Tilde{E}_{s}(k_x,k_y)\mathbf{e}_s(k_x,k_y)e^{-j\mathbf{k.r}}dk_xdk_y \\
    &+ \frac{1}{(2\pi)^2}\int\int \Tilde{E}_{p}(k_x,k_y)\mathbf{e}_p(k_x,k_y)e^{-j\mathbf{k.r}}dk_xdk_y
    \end{split}
\end{equation}
where $\Tilde{E}_{s}(k_x,k_y)$ and $\Tilde{E}_{p}(k_x,k_y)$ are complex amplitudes of the $TE_z$ (s-polarized) and $TM_z$ (p-polarized) plane waves propagating in the medium along corresponding wavevector $\mathbf{k}=(k_x,k_y,k_z)$ with $k=|\mathbf{k}|=n\frac{\omega}{c}$. It is also noted that $\mathbf{e}_s(k_x,k_y)=(-k_y/k_t,k_x/k_t,0)$ and $\mathbf{e}_p(k_x,k_y)=(-k_xk_z/kk_t,-k_yk_z/kk_t,k_t/k)$ are the corresponding unit vectors of $TE_z$ and $TM_z$ electric field where $k_t=|\mathbf{k_t}|$ and $\mathbf{k_t}=k_x\hat{x}+k_y\hat{y}$ is the transverse wavevector.
Plane wave amplitudes $\Tilde{E}_{s}(k_x,k_y)$ and $\Tilde{E}_{p}(k_x,k_y)$ can be properly calculated by Fourier expansion of any two components of the electromagnetic field at an arbitrary transverse plane $z=cte$. In other words, the spatial frequency content of $\mathbf{E}(x,y,z=cte)$, mathematically corresponding to the spatial frequency pair of $(k_x,k_y)$, is actually the set of amplitudes of plane waves that together form the beam.

Performing linear mathematical operations on an input field profile can be described by properly manipulating the corresponding frequency content or equivalently, the amplitudes of plane waves in the wavevector domain. A mathematical description of such an operation is addressed by an Optical Transfer Function (OTF) as:
\begin{equation} \label{equ:OTF}
    \underline{H}(k_x,k_y)=
    \begin{pmatrix}
        H_{ss}(k_x,k_y)&H_{ps}(k_x,k_y)\\
        H_{sp}(k_x,k_y)&H_{pp}(k_x,k_y)\\
    \end{pmatrix}
\end{equation}
The diagonal entries of an OTF describe the actual transfer functions applied on TE and TM polarizations and for conventional design approaches must ideally be identical. On the other hand, the off-diagonal entries indicate possible couplings between polarizations and should ideally vanish. Nevertheless, some optical operators are intended to function on cross-polarization mode i.e. the desired functionality is embedded in the off-diagonal entries of OTF \cite{Zhu2019}. Applying an OTF on input plane waves results in a set of output plane waves that can be obtained straightforwardly as:
\begin{equation} \label{equ:OTFapplication}
    \begin{pmatrix}
        \Tilde{E}_{s}^{o}(k_x,k_y)\\
        \Tilde{E}_{p}^{o}(k_x,k_y)\\
    \end{pmatrix}
    =\underline{H}(k_x,k_y)
    \begin{pmatrix}
        \Tilde{E}_{s}^{i}(k_x,k_y)\\
        \Tilde{E}_{p}^{i}(k_x,k_y)\\
    \end{pmatrix}
\end{equation}
The combination of these output plane waves on a so-called output plane forms the output wavefront. Thus, the numerical demonstrations for performing an optical analog operator on an input wavefront can be accomplished by applying the corresponding OTF on input plane wave amplitudes and then integrating over output plane waves using the inverse Fourier transform of \eqref{equ:FourierDecomp}. This procedure is used in the upcoming sections along with Finite Element Method (FEM) simulations to verify the obtained results and to show the application of the proposed structures.

In this paper, our objective is to design an optical device to implement high pass filtering on input beam profiles in the wavevector domain. High pass filtering is a mathematical linear operation that blocks low-frequency spectral content of an input signal and in an ideal case, uniformly passes the other part of spectral content with frequencies higher than a specific threshold, the cut-off frequency $k_c$, as follows:
\begin{equation} \label{equ:HPF}
H(k_t)=\begin{cases} 0 & |k_t|<k_c \\
                     1 & |k_t| \geq k_c
       \end{cases}
\end{equation}
It is noteworthy that the transfer function of an ideal HPF in \eqref{equ:HPF} depends only on the magnitude of transverse wavevector and not its direction, hence "isotropic". High pass filters with isotropic OTF may be utilized for resolving sharp edges in 2D input profiles uniformly along any direction. Other designs have also been proposed in the literature for 2D edge detection, using anisotropic \cite{Saba2018} and nearly isotropic \cite{Guo2018:diff} Laplacian operators.

\section{Design, Results and Discussion}
In the following subsections, by exploiting the formalism developed in previous section, we design and analyze two optical structures for implementing 1D and 2D spatial high pass filtering to perform edge detection of input field profiles as an application.

\subsection{1D High Pass Filter}
\begin{figure}[t]
    \centering
\begin{subfigure}[b]{0.3\textwidth}
         \centering
         \includegraphics[width=\textwidth]{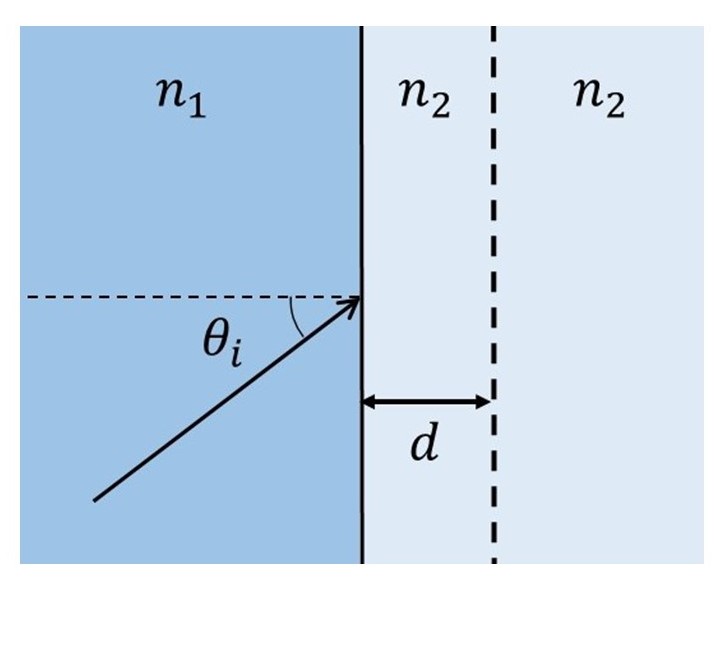}
         \caption{}
         \label{fig:1DHPFstructure(a)}
     \end{subfigure}
     \hfill
     \begin{subfigure}[b]{0.31\textwidth}
         \centering
         \includegraphics[width=\textwidth]{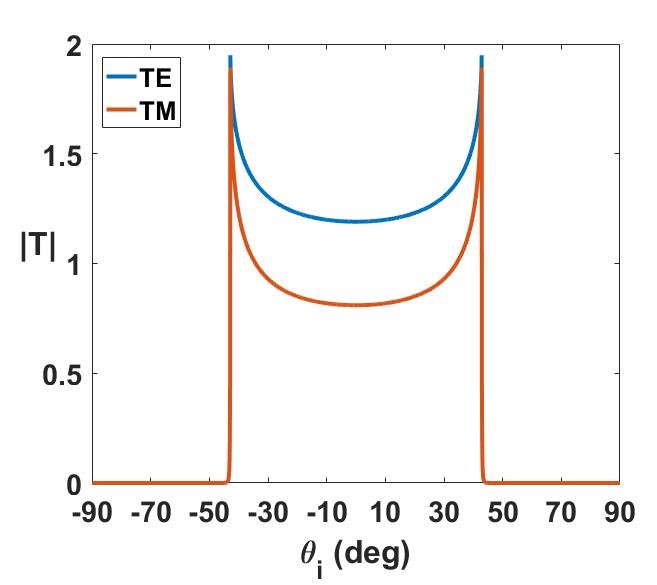}
         \caption{}
         \label{fig:1DHPFstructure(b)}
     \end{subfigure}
     \hfill
     \begin{subfigure}[b]{0.31\textwidth}
         \centering
         \includegraphics[width=\textwidth]{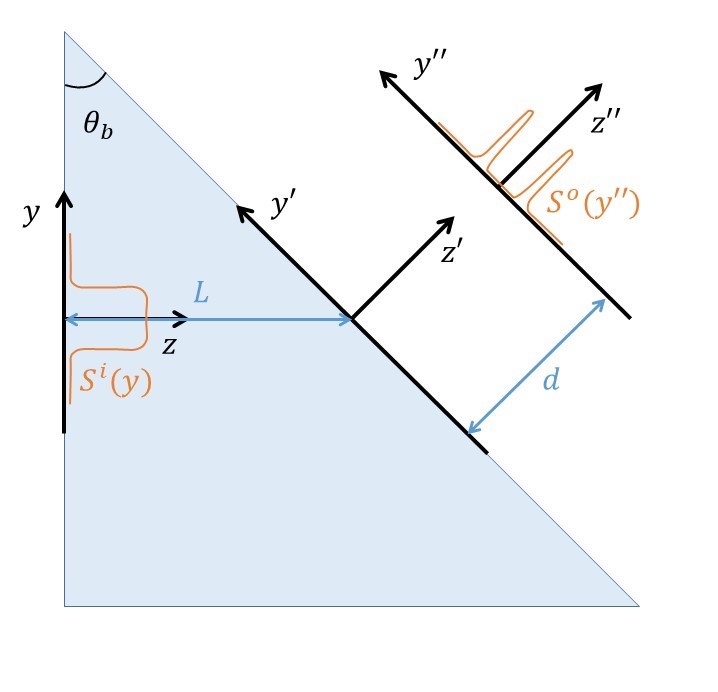}
         \caption{}
         \label{fig:1DHPFstructure(c)}
     \end{subfigure}
        \caption{1D HPF structure: (a) The interface of two dielectric media and (b) its corresponding transmission coefficient for $n_1=1.47$, $n_2=1$ and $d=5\lambda_0$. (c) The proposed structure for implementing a 1D HPF based on total internal reflection using a prism}
        \label{fig:1DHPFstructure}
\end{figure}

The reflection coefficients of plane waves from a planar interface between two dielectric media with different refractive indices of $n_1$ and $n_2$ as shown in Fig.~\ref{fig:1DHPFstructure(a)} are calculated from Fresnel equations for TE and TM polarizations as follows:
\begin{equation} \label{equ:rFresnel}
    \begin{aligned}
    R_{TE} &= \frac{n_1cos\theta_i-\sqrt{n_2^2-n_1^2sin^2\theta_i}}{n_1cos\theta_i+\sqrt{n_2^2-n_1^2sin^2\theta_i}}\\        
    R_{TM} &= \frac{n_2^2cos\theta_i-n_1\sqrt{n_2^2-n_1^2sin^2\theta_i}}{n_2^2cos\theta_i+n_1\sqrt{n_2^2-n_1^2sin^2\theta_i}}
  \end{aligned}
\end{equation}
Assuming $n_1>n_2$, the reflection coefficients have complex values with unit amplitudes for incident angles greater than the critical angle  $\theta_c=sin^{-1}({n_2}/{n_1})$. This means that for such angles, the incident plane waves from medium 1 are non-propagating in medium 2 and will be evanescent along axis that is normal to the interface and thus no power will be transmitted to the other side of the interface by these plane waves. Assuming the output plane to be located at a distance $d$ from the interface in medium 2 as shown in Fig.~\ref{fig:1DHPFstructure(a)}, the transmission coefficients of incident plane waves are calculated as a function of incident angle for TE and TM polarizations. Fig.~\ref{fig:1DHPFstructure(b)} shows the amplitude of transmission coefficients for $n_1=1.47$, $n_2=1$ and $d=5\lambda_0$. As one can deduce from Fig.~\ref{fig:1DHPFstructure(b)} the critical angle for this structure is $\theta_c=45.6^{\circ}$ and greater incident angles face total internal reflection (TIR).

To use the TIR effect for spatial high pass filtering, the low-frequency content of the input wavefront should coincide with the angular section of totally reflected wavevectors and be absent on the output plane. This is possible by a translation in the wavevector domain which is equivalent to a rotation in the spatial domain. For this purpose, the input beam profile should illuminate the interface at an oblique angle $\theta_b$ that is slightly greater than the critical angle $\theta_b>\theta_c$. The proposed configuration for this functionality is illustrated in Fig.~\ref{fig:1DHPFstructure(c)} where a single prism is shown to serve as a 1D high pass filter. It must be noted that this structure provides an asymmetric (single-sided) high pass filtering performance. This is because only one side of the high-frequency content ($k_y>n_1k_0sin(\theta_b-\theta_c)$ with respect to notation of Fig.~\ref{fig:1DHPFstructure(c)}) will pass through the interface whereas the other side (negative values of $k_y$) will be totally reflected. It is noteworthy that the parameter ($\theta_b-\theta_c$) determines the width of stopband of the filter.

\begin{figure}[t]
     \centering
     \begin{subfigure}[b]{0.49\textwidth}
         \centering
         \includegraphics[width=\textwidth]{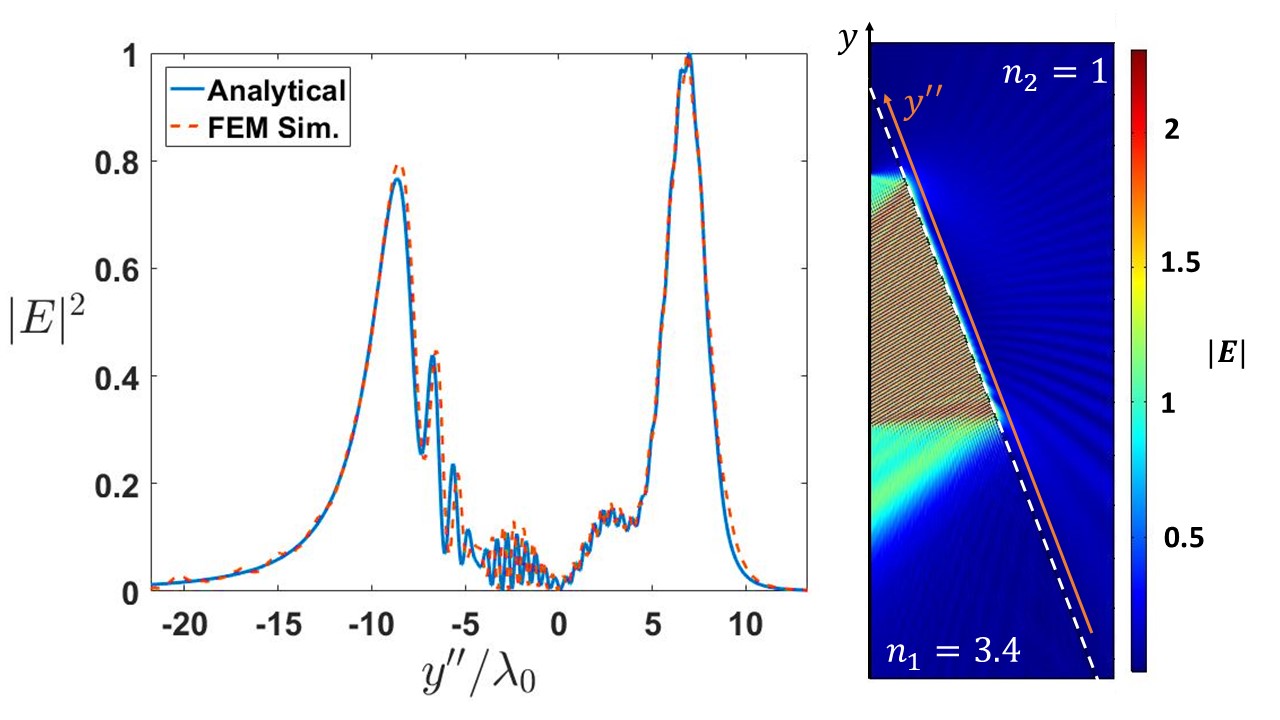}
         \caption{}
         \label{fig:1DHPFslitedge(a)}
     \end{subfigure}
     \hfill
     \begin{subfigure}[b]{0.49\textwidth}
         \centering
         \includegraphics[width=\textwidth]{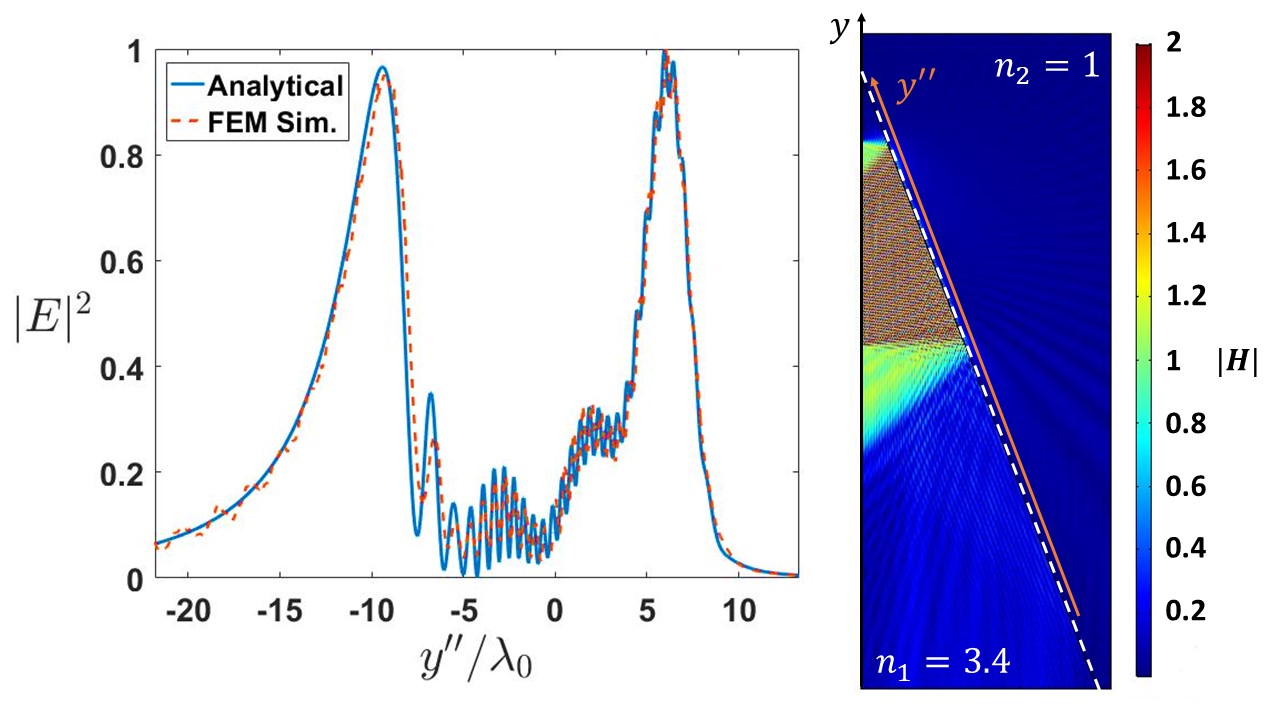}
         \caption{}
         \label{fig:1DHPFslitedge(b)}
     \end{subfigure}
     \begin{subfigure}[b]{0.3\textwidth}
         \centering
         \includegraphics[width=\textwidth]{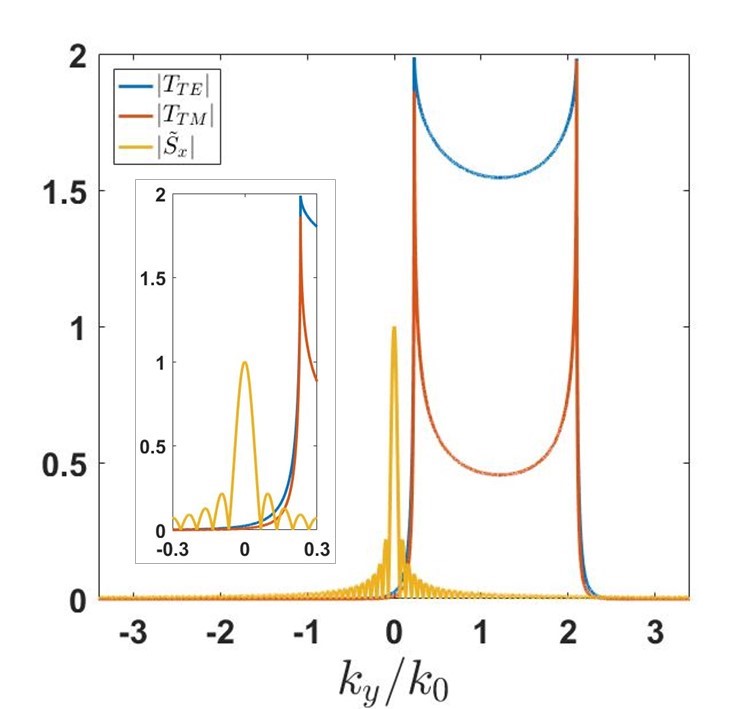}
         \caption{}
         \label{fig:1DHPFslitedge(c)}
     \end{subfigure}
     \hfill
     \begin{subfigure}[b]{0.34\textwidth}
         \centering
         \includegraphics[width=\textwidth]{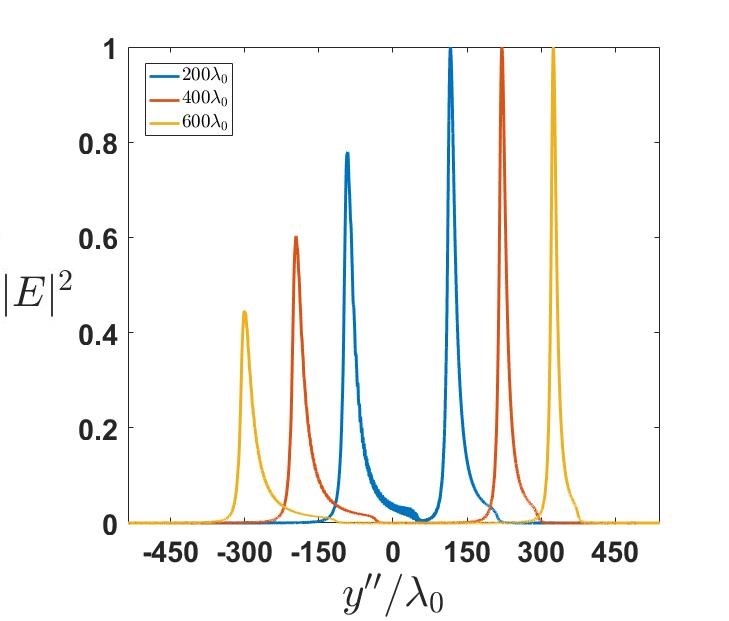}
         \caption{}
         \label{fig:1DHPFslitedge(d)}
     \end{subfigure}
     \hfill
     \begin{subfigure}[b]{0.34\textwidth}
         \centering
         \includegraphics[width=\textwidth]{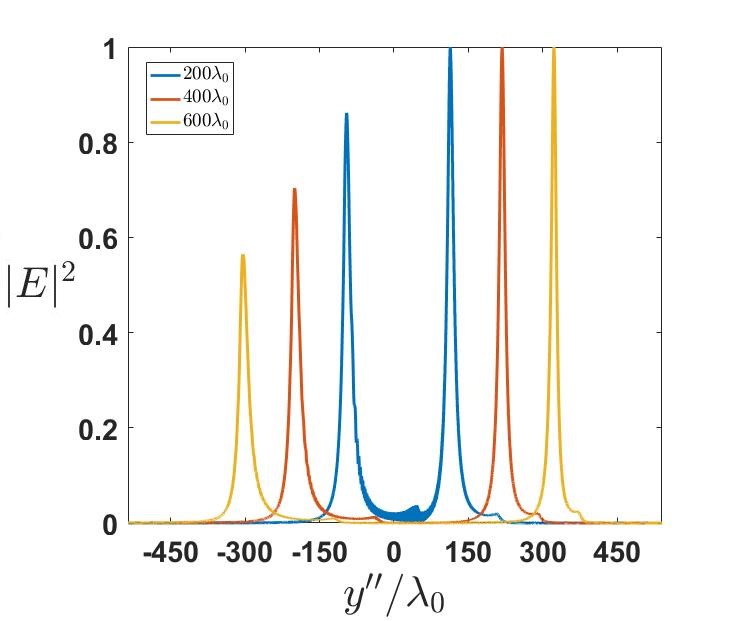}
         \caption{}
         \label{fig:1DHPFslitedge(e)}
     \end{subfigure}
        \caption{Analytical and numerical results for 1D HPF using a silicon prism with $n_1=3.4$ and air $n_2=1$ with free space propagation length of $d=\lambda_0$ applied on slit width of $15\lambda_0$ for (a) TE (b) and TM polarized illumination. The field distribution is shown at right hand side of (a) and (b) where a rectangular aperture illuminates the interface (white dashed line located at diagonal) from left in accordance with configuration of Fig.~\ref{fig:1DHPFstructure(c)}. (c) Corresponding wavevector domain representation of input fields and the applied OTF for each polarization. Abrupt variation in the input rectangular apertures with different widths have transformed into sharp peaks in the output field profile for both (d) TE and (e) TM.}
        \label{fig:1DHPFslitedge}
\end{figure}

For a more quantitative analysis, we assume a 1D input profile $S^i(y)$ on the y axis as shown in Fig.~\ref{fig:1DHPFstructure(c)} which represents the $x$ component of the electric (magnetic) field for $TE_z (TM_z)$ polarization. Fourier expansion determines the complex plane wave amplitude of the input profile as $\Tilde{S}^i(k_y)$. Substituting $y$ and $z$ in terms of $y'$ and $z'$ in \eqref{equ:FourierDecomp} to rewrite plane wave amplitudes in the primed coordinates and then applying the OTF of transmission coefficient $T(k_{y'})$ calculated as a function of $k_{y'}$, the output profile is obtained on the output plane by integrating over all of the plane waves as follows:
\begin{equation} \label{equ:1DHPFintegral}
    S^{o}(y'')=\frac{1}{2\pi}\int T(k_{y'})\Tilde{S}^i(k_y)e^{-jk_{y'}y''}e^{-jk_zL}dk_y
\end{equation}
where $k_{y'}=k_ycos\theta_b+k_zsin\theta_b$. Note that since the variations of the input wavefront have been assumed only along $y$ axis, the $x$ component of wavevectors vanishes. It is a good approximation for a 2D paraxial input profile as well. The $e^{-jk_zL}$ factor in \eqref{equ:1DHPFintegral} accounts for the phase accumulation of plane waves due to propagating along $z$ axis.

In the following, the interface of a silicon prism, $n_1=3.4$, and air, $n_2=1$, with free space propagation length of $d=\lambda_0$ is assumed to implement the desired HPF for an illuminated rectangular apertures with different slit widths as $S^i(y)$. The incident beam angle has been considered to be $\theta_b=20^{\circ}$ which is slightly greater than $\theta_c=17.1^{\circ}$. Figures \ref{fig:1DHPFslitedge(a)} and \ref{fig:1DHPFslitedge(b)} illustrate the numerical simulation of the proposed structure (using FEM) in comparison to analytical results obtained from \eqref{equ:1DHPFintegral} for a slit width of $15\lambda_0$ with TE and TM polarizations, respectively. Wavevector domain representation of corresponding input fields and the applied OTF for each polarization are also illustrated in Fig.~\ref{fig:1DHPFslitedge(c)}. It is observed that numerical and analytical results are in good agreement and thus the design is verified. Applying \eqref{equ:1DHPFintegral} to calculate the wavefront on the output plane for wider slit widths with $d=2\lambda_0$ and $\theta_b=21^{\circ}$ produces the results shown in Fig.~\ref{fig:1DHPFslitedge(d)} and Fig.~\ref{fig:1DHPFslitedge(e)} for TE and TM polarization, respectively. Notably, the performance of the proposed spatial high pass filter is not dependent on the polarization of the input field and a single device provides the functionality of interest for both TE and TM illumination and therefore unpolarized light. This originates from the independence of the critical angle into polarization to take place the TIR effect. It should also be emphasized that the non-resonant nature of TIR facilitates the reconfigurability of the stopband of the filter, that in turn enables the high-resolution detection of sharp edges in the input profile. The independence of TIR from incident wavelength, given a negligible material dispersion for dielectrics, offers a wideband functionality in the whole visible range.

\begin{figure}[t]
    \centering
\begin{subfigure}[b]{0.32\textwidth}
         \centering
         \includegraphics[width=\textwidth]{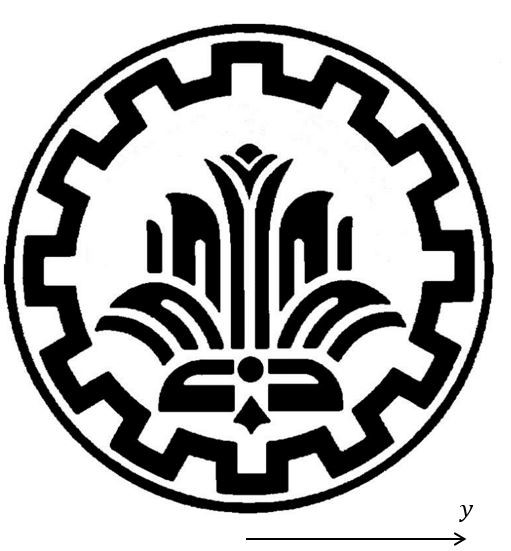}
         \caption{}
         \label{fig:SharifLogoEdge_input}
     \end{subfigure}
     \hfill
     \begin{subfigure}[b]{0.33\textwidth}
         \centering
         \includegraphics[width=\textwidth]{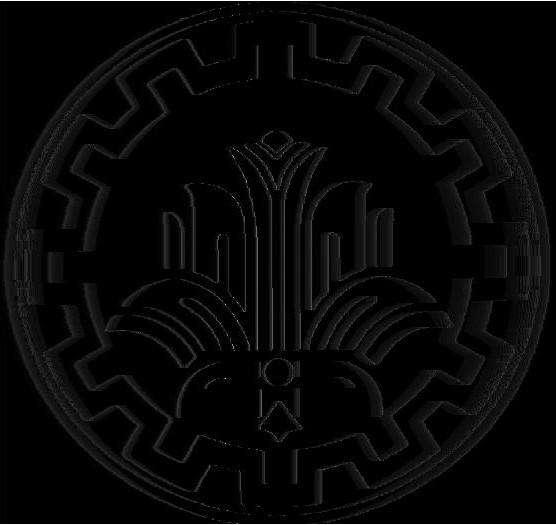}
         \caption{}
         \label{fig:SharifLogoEdge_output_TE}
     \end{subfigure}
     \hfill
     \begin{subfigure}[b]{0.33\textwidth}
         \centering
         \includegraphics[width=\textwidth]{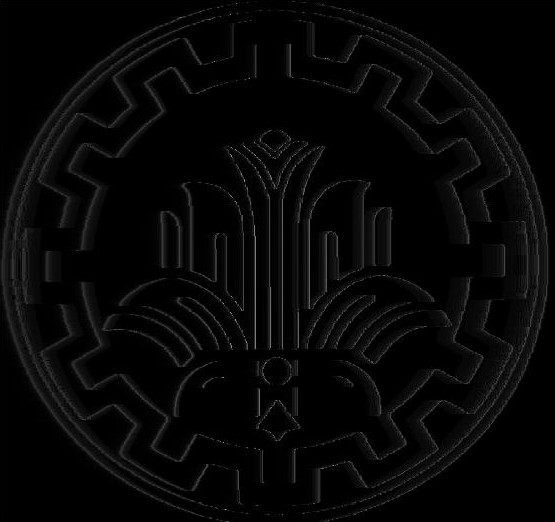}
         \caption{}
         \label{fig:SharifLogoEdge_output_TM}
     \end{subfigure}
        \caption{1D edge detection (a) Logo of Sharif University of Technology as input profile and corresponding y axis that the 1D high pass filtering has been applied along it for input profile size of $7420\lambda_0\times7420\lambda_0$ and corresponding (b) TE and (c) TM output wavefronts.}
        \label{fig:SharifLogoEdge}
\end{figure}

We can now demonstrate the application of the designed optical device for practical purposes. Detecting edges of objects in images is an essential part of object recognition and feature extraction processes. Performing this step in an analog fashion provides a computational advantage in terms of computation power and time over digital algorithms especially in case of a large amount of input data. Fig.~\ref{fig:SharifLogoEdge} demonstrates the operation of our 1D HPF for edge detection in which the logo of Sharif University of Technology is used as input. Figures \ref{fig:SharifLogoEdge_output_TE} and \ref{fig:SharifLogoEdge_output_TM} are respectively the output of TE and TM polarized illumination of input. It is seen that sharp spatial variations along $y$ axis in the input field transform into bright border lines right at the location of edges in the output plane. However, oblique edges that have projections in both $x$ and $y$ axes have been detected as well. These results have been obtained by numerically implementing the integral of \eqref{equ:1DHPFintegral} and substituting the TE or TM transfer function for $T(k_{y'})$. Moreover, the center of input profile in Fig.~\ref{fig:SharifLogoEdge_input} has been focused on the interface in our calculations.

\subsection{2D High Pass Filter}
\begin{figure}[t]
    \centering
    \begin{subfigure}[b]{0.33\textwidth}
        \centering
        \includegraphics[width=\textwidth]{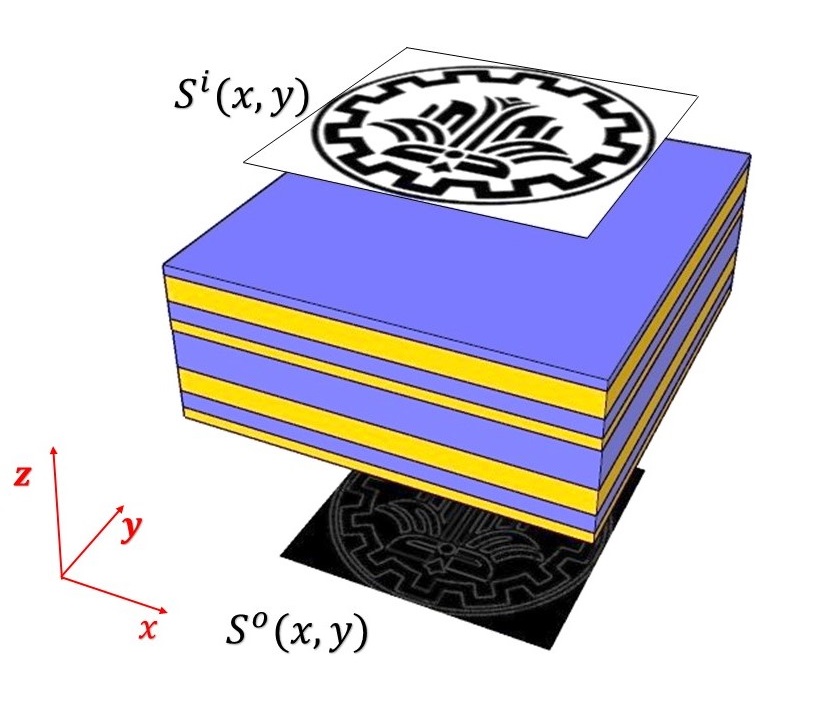}
        \caption{}
        \label{fig:2D_HPF_structure}
    \end{subfigure}
    \hfill
    \begin{subfigure}[b]{0.33\textwidth}
         \centering
         \includegraphics[width=\textwidth]{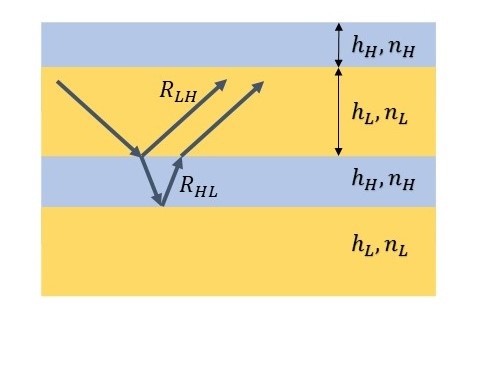}
        \caption{}
        \label{fig:Bragg_partial_reflection}
     \end{subfigure}
     \hfill
     \begin{subfigure}[b]{0.33\textwidth}
         \centering
         \includegraphics[width=\textwidth]{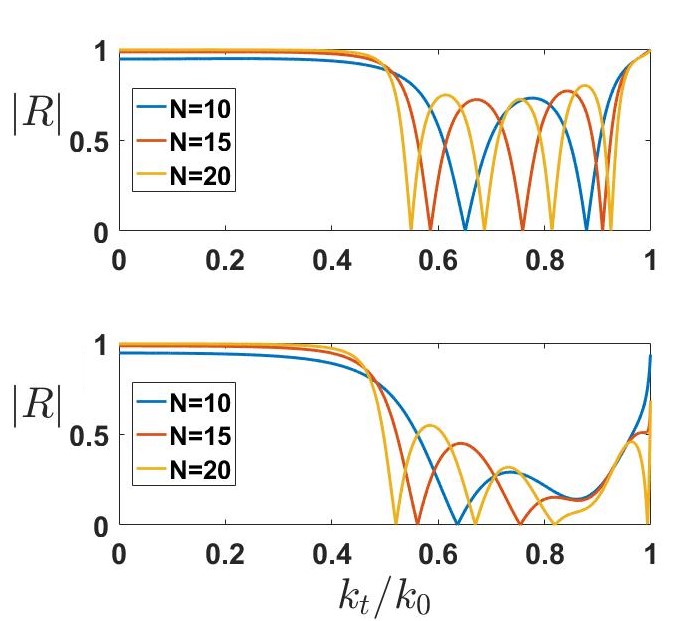}
         \caption{}
         \label{fig:QW_differentN}
     \end{subfigure}
        \caption{(a) Schematics of multilayer dielectric structure to implement isotropic 2D spatial high pass filter. (b) Partial reflections from boundaries $R_{HL}$ and $R_{LH}$ are only different in a minus sign i.e. a difference of $\pi$ in phase. By using quarter-wave layers for another phase of $\pi$ to be accumulated in a round-trip \eqref{equ:Bragg_condition}, constructive interference of reflected waves shall occur. (c) TE (up) and TM (down) Reflection coefficients of quarter-wave layers of dielectrics with $n_H=1.7$ and $n_L=1.5$ for different numbers of unit cell repetition $N$. It is shown that the stopband does not depend on $N$.}
        \label{fig:2DHPF}
\end{figure}

In this subsection, we use the transmission coefficient of a dielectric multilayer structure to implement a 2D spatial high pass filter in the wavevector domain. Therefore, the low-frequency content of the input field ,$|k_t|<k_c$ in \eqref{equ:HPF}, should be reflected and fade out on the transmission side. The proposed structure for this purpose is shown in Fig.~\ref{fig:2D_HPF_structure} where an input beam profile $\mathbf{S}^i(x,y)$ illuminates a stack of dielectric layers with a normal incident angle. It is noteworthy that according to the symmetry of the structure with respect to the beam axis, the OTF of interest will readily be isotropic i.e. $\underline{H}(k_x,k_y)=\underline{H}(k_t)$. To have high pass filtering performance, the constituent plane waves with small transverse wavevectors or equivalently near normal incident angles must be blocked. In the following, we start the design of a 2D HPF using Bragg condition on a periodic multilayer stack of dielectrics. Then, we tune the periodicity of the structure to adjust the stopband of the transfer function. Finally, by minimizing a \textit{cost function} via optimizing the thickness of layers, the designed OTF will emulate a closer approximation of the desired transfer function of \eqref{equ:HPF}.

Our goals to implement a 2D HPF functionality are set as follows: (i) to keep the reflection coefficient as close to unity as possible for normal incidence, (ii) to keep the stopband relatively narrow so that the output for paraxial input profiles not be immersed in noise due to weak transmitted signal and (iii) to have the diagonal elements of OTF in \eqref{equ:OTF} manifest almost identical responses. The first item (i) in the list requires a noticeable contrast between refractive indices $n_H$ and $n_L$ which opposes the requirements of the other two items.

To maximize reflection from a periodic multilayer structure, the thickness of layers in each period should be chosen in such a way that the partial reflections from boundaries interfere constructively in the incidence medium (Fig.~\ref{fig:Bragg_partial_reflection}). This condition is called Bragg reflection condition and for a periodic structure with two layers in each period (Layers L and H) imposes quarter-wave layers as denoted in following for normal incidence: 
\begin{equation} \label{equ:Bragg_condition}
2n_ik_0h_i=\pi \quad \Rightarrow \quad n_ih_i=\frac{\lambda_0}{4} \quad ; \quad i=L,H 
\end{equation}
where $k_0$ and $\lambda_0$ are wave-number and wavelength of free space, respectively. $n_i$ and $h_i$, $i=L,H$ determine refractive index and thickness of each layer. The total reflection coefficient $\Gamma$ of such a periodic structure at a normal incident angle is calculated by transforming the impedance of each layer starting from the last boundary as follows:
\begin{equation} \label{equ:Gamma_reflection}
    Z_1=\frac{\eta_H^2}{Z_2}=\frac{\eta_H^2}{\eta_L^2}Z_3=\eta_0(\frac{\eta_H}{\eta_L})^{2N} \quad \Rightarrow \quad \Gamma=\frac{Z_1-\eta_0}{Z_1+\eta_0}=\frac{1-(\frac{n_H}{n_L})^{2N}}{1+(\frac{n_H}{n_L})^{2N}} 
\end{equation}
where $Z_j$ is the wave impedance at $j$th boundary, $N$ is the repetition number of unit cells and we have assumed that the incident and exit media are air i.e. $n_{inc}=n_{ext}=1$. This reflection coefficient approaches unity for larger $N$ and is actually the maximum possible reflection coefficient attained for certain values of $N$, $n_H$ and $n_L$. This will play a major role in our HPF design since by optimizing the layer thicknesses, the reflection coefficient will deviate from its maximum value at normal incidence. Therefore, we have to leave enough room for this deviation by choosing appropriate values of $N$, $n_H$ and $n_L$ so that the final optimized structure can still exhibit a reflection of high enough modulus at near normal angles.

Reflection and transmission coefficients of plane waves incident on a stack of planar layers with different layer thicknesses and refractive indices can easily be calculated using the transfer matrix method (TMM) \cite{TMMAnemogiannis1999}. Fig.~\ref{fig:QW_differentN} shows the total reflection coefficient of a quarter-wave stack of dielectrics with $n_H=1.7$ and $n_L=1.5$ for both polarizations and different values of $N$. As shown in Fig.~\ref{fig:QW_differentN}, the structure exhibits a reflective characteristic in a range of wavevectors that we call it the stopband and an oscillatory transmission is observed for other wavevectors. Evidently, increasing $N$ does not change the stopband width but causes the transmission to be more oscillatory in the passband. In the following, we investigate the design parameters of a periodic dielectric stack to engineer the stopband width. Our analysis is based on the transfer matrix properties of a unit cell \cite{Orfanidis}.

\begin{figure}[t]
    \centering
    \begin{subfigure}[b]{0.32\textwidth}
        \centering
        \includegraphics[width=\textwidth]{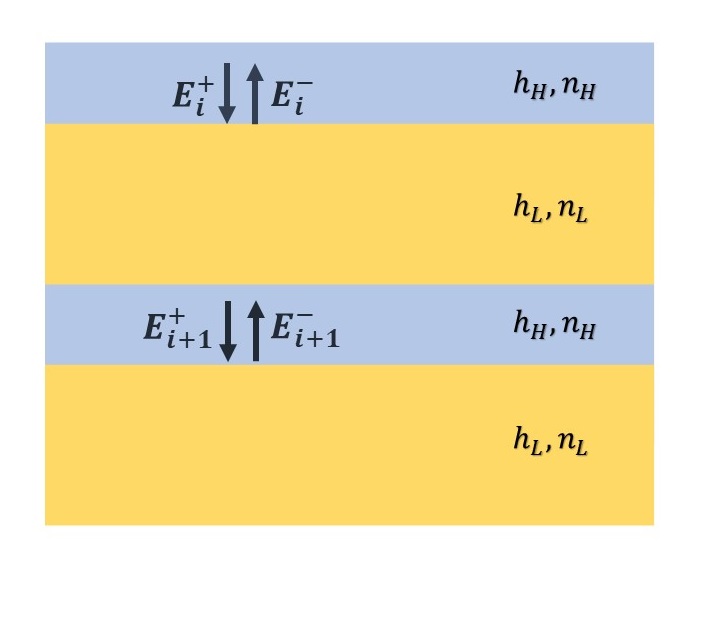}
        \caption{}
        \label{fig:unitcell_matrix}
    \end{subfigure}
    \hfill 
    \begin{subfigure}[b]{0.33\textwidth}
        \centering
        \includegraphics[width=\textwidth]{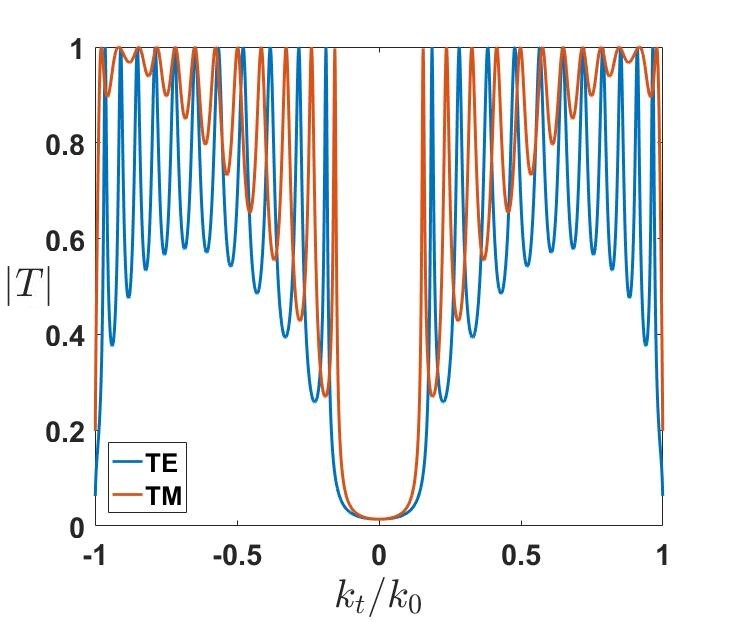}
        \caption{}
        \label{fig:2D_HPF_startpoint}
    \end{subfigure}
    \hfill
    \begin{subfigure}[b]{0.34\textwidth}
         \centering
         \includegraphics[width=\textwidth]{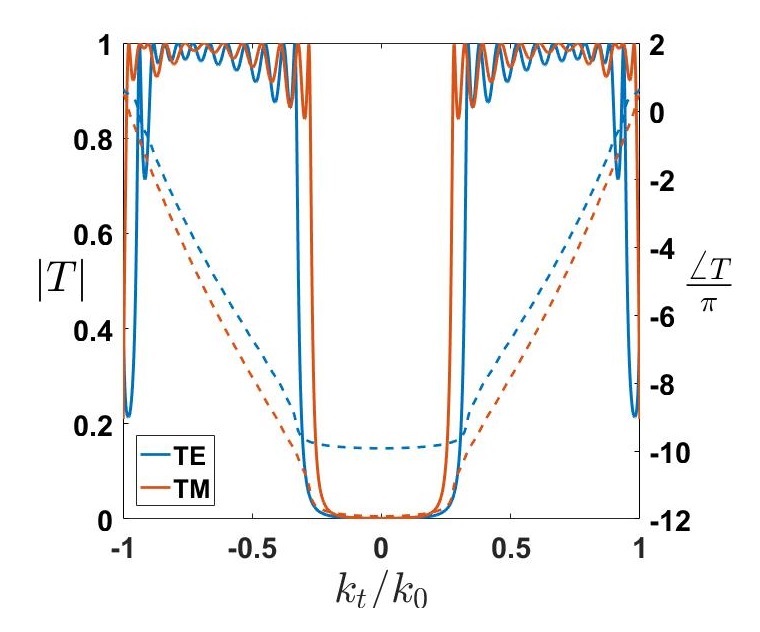}
        \caption{}
        \label{fig:optim_2DHPF}
     \end{subfigure}
    \caption{(a) Forward and backward field amplitudes at the topmost interface of two consequent unit cells that are related via a transfer matrix. (b) The transmission coefficient of plane waves incident with transverse wavevector $k_t$ on a periodic stack of dielectrics with $n_ih_i=\frac{\lambda_0}{4.37}$ and (c) its optimized version to implement 2D high pass filter where solid lines show the amplitude and dashed lines show the phase of transmission for both polarizations.}
    \label{fig:2DHPF_results}
\end{figure}

The forward and backward field amplitudes at the topmost interface of each unit cell in a periodic stack i.e. a 1D photonic crystal as depicted in Fig.~\ref{fig:unitcell_matrix}, can be related to the forward and backward field amplitudes of the next unit cell via a transfer matrix $F$. The eigenvalues of $F$ can be shown to be of the form $\lambda _{\pm}=e^{\pm jKL}$ and determine the characteristic of the crystal to be reflective or transmissive. Here $L=h_H+h_L$ is the periodicity and $K$ is known as the Bloch wavevector of the structure. Depending on the design parameters, incident angle and polarization, values of $K$ may be real, in which case $\lambda _{\pm}$ will become pure phase factors resulting in oscillatory transmission in the passband, or imaginary that yield descending exponential factors. In the latter case, the incident plane waves will propagate evanescently normal to the stack of layers and  be reflected almost completely for a high enough number of unit cells $N$.

Choosing $n_H=2$ (SiN) and $n_L=1.52$ (BK7) and $N=50$ and also tailoring the periodicity to have $n_ih_i=\frac{\lambda_0}{4.37}$, our design can partially enjoy the qualities of a 2D optical HPF in transmission mode as depicted in Fig.~\ref{fig:2D_HPF_startpoint}. The resulted OTF has a relatively acceptable performance in the stopband but the passband functionality is highly oscillatory rather than being uniform and smooth for a proper HPF. Finally, an optimization algorithm shall be used to improve our design and alleviate the shortcomings in the passband functionality. To formulate the problem in terms of an optimization problem, we define a \textit{cost function} as: 
\begin{equation} \label{equ:costfunction}
    c.f.=\sum_{k_t=0}^{k_0} \frac{|H_{target}(k_t)-|T(k_t)||}{w(k_t)}
\end{equation}
that evaluates a weighted distance between the realized OTF and the target OTF of \eqref{equ:HPF}. Thus, the problem reduces to minimizing the so-called cost function by adjusting the layer thicknesses. This problem is tackled by using the \textit{fminsearch} function in MATLAB\texttrademark{} that runs a simplex algorithm to seek for the local minimum of c.f. in the vicinity of the starting point. The weight function $w(k_t)$ introduces the acceptable tolerance in different parts of the OTF to our optimization. Figure \ref{fig:optim_2DHPF} demonstrates the amplitude and phase of the transmission coefficient of the realized OTF after optimizing the layer thicknesses in our multilayer slab for both TE and TM polarizations, starting from OTF shown in Fig.~\ref{fig:2D_HPF_startpoint}. This provides us the desired characteristic of a polarization-insensitive 2D spatial HPF. It is seen in Fig.~\ref{fig:2D_HPF_startpoint} that the overall transmission coefficient modulus for TM polarized plane waves approaches unity for grazing angles of incidence and is less oscillatory than the case of TE polarization. It is thus convincing that optimizing only the TE-polarized OTF will also satisfy the requirements of the TM-polarized OTF, as is the case here.

\begin{figure}[t]
     \centering
          \begin{subfigure}[b]{0.32\textwidth}
         \centering
         \includegraphics[width=\textwidth]{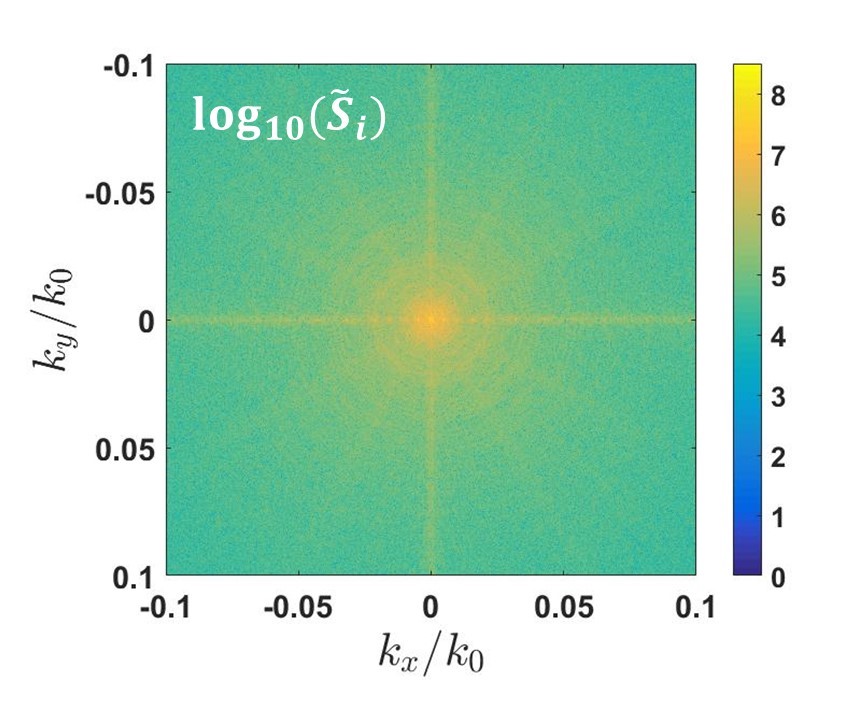}
         \caption{}
         \label{fig:2D_HPF_logo_edge(a)}
     \end{subfigure}
     \hfill
     \begin{subfigure}[b]{0.33\textwidth}
         \centering
         \includegraphics[width=\textwidth]{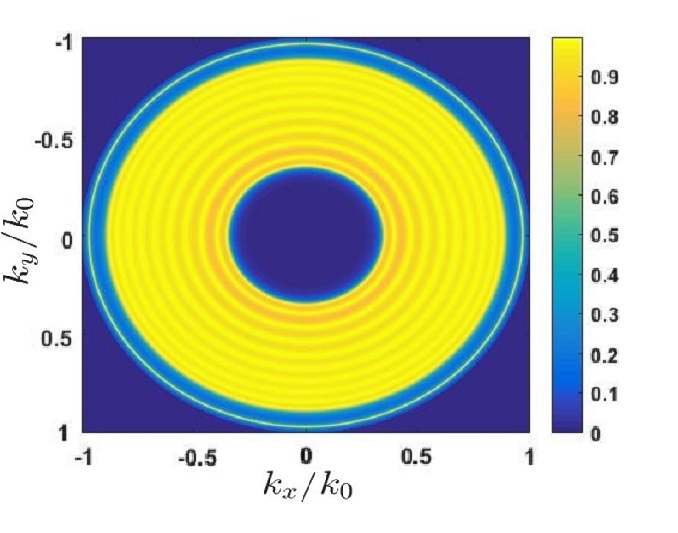}
         \caption{}
         \label{fig:2D_HPF_logo_edge(b)}
     \end{subfigure}
     \hfill
     \begin{subfigure}[b]{0.33\textwidth}
         \centering
         \includegraphics[width=\textwidth]{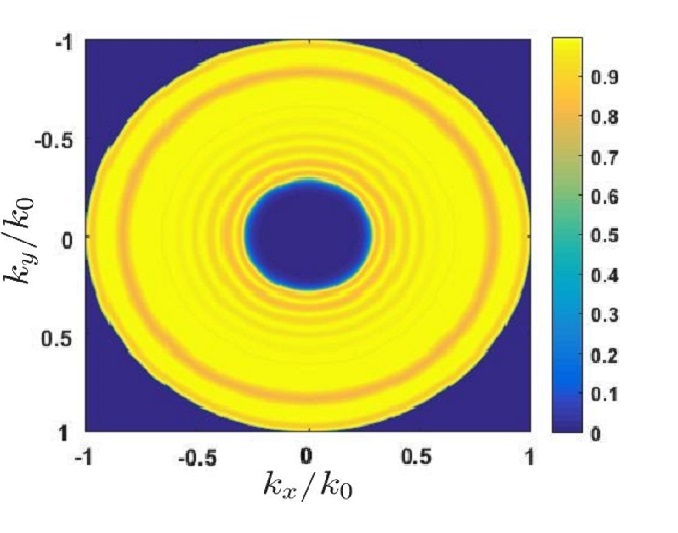}
         \caption{}
         \label{fig:2D_HPF_logo_edge(c)}
     \end{subfigure}
     \\[3ex] 
     \begin{subfigure}[b]{0.44\textwidth}
         \centering
         \includegraphics[width=\textwidth]{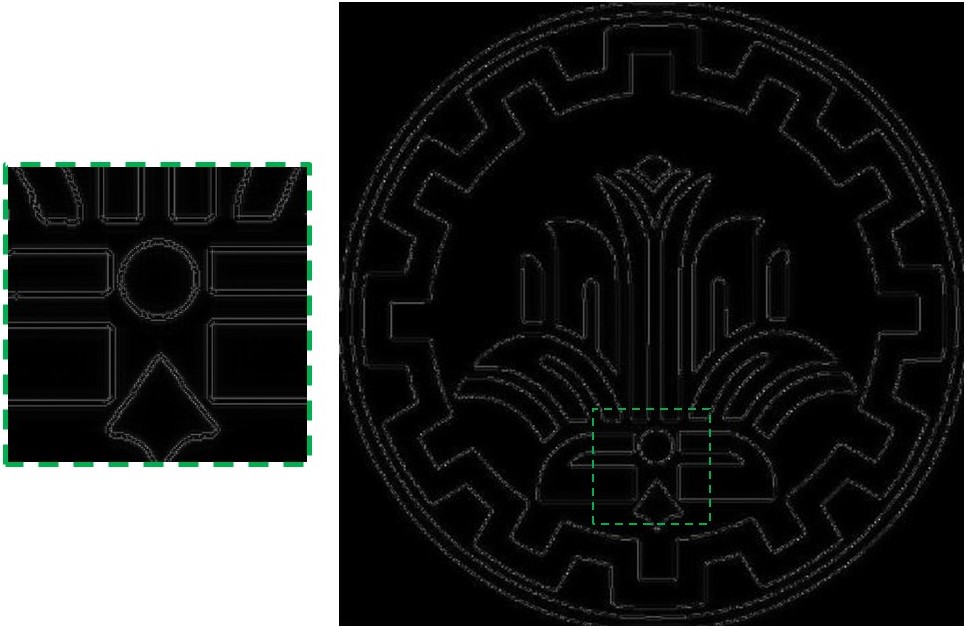}
         \caption{}
         \label{fig:2D_HPF_logo_edge(d)}
     \end{subfigure}
     \quad
     \begin{subfigure}[b]{0.44\textwidth}
         \centering
         \includegraphics[width=\textwidth]{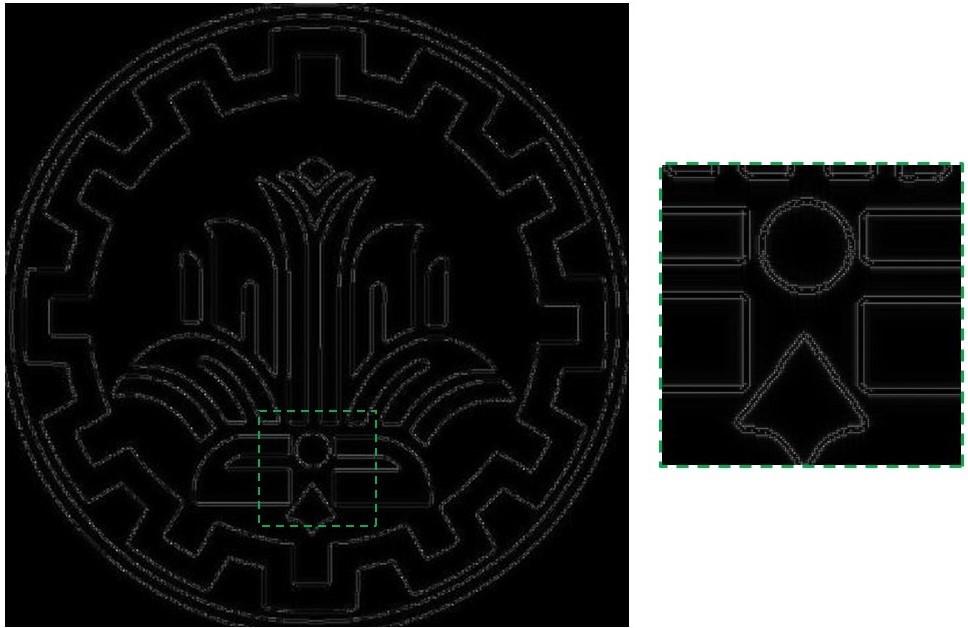}
         \caption{}
         \label{fig:2D_HPF_logo_edge(e)}
     \end{subfigure}
    \caption{2D edge detection: (a) Representation of SUT logo in wavevector domain as input profile with input size of $3710\lambda_0\times3710\lambda_0$. The isotropic optimized OTF of multilayer dielectric structure for (b) TE and (c) and TM polarizations.  Corresponding output wavefronts for (d) TE and (e) TM polarizations.}
    \label{fig:2D_HPF_logo_edge}
\end{figure}

Applying the designed optical filter for detecting edges of the 2D input profile shown in Fig.~\ref{fig:SharifLogoEdge_input} finishes our discussion on performing spatial high pass filtering in an analog fashion. Figures \ref{fig:2D_HPF_logo_edge(a)}, \ref{fig:2D_HPF_logo_edge(b)} and \ref{fig:2D_HPF_logo_edge(c)} show the corresponding representation of the input profile in the wavevector domain as well as a 2D illustration of the applied filters for TE and TM polarizations. Our filter is completely isotropic due to transversal symmetry. The transmitted plane waves will eventually combine on the other side of the slab to form the output wavefront. It is calculated via inverse Fourier transform of \eqref{equ:FourierDecomp} that has been carried out numerically to produce the output results of Fig.~\ref{fig:2D_HPF_logo_edge(d)} and Fig.~\ref{fig:2D_HPF_logo_edge(e)} in our case. Noteworthy, a uniform detection of edges regardless of their direction is evident. Moreover, the negative and positive slope of the applied phase on both sides of the passband, depicted in Fig.~\ref{fig:optim_2DHPF}, cause the highlighted edges in the output profile to experience two undesirable shifts in the spatial domain. However, such a slight shift can be neglected for practical applications where the distance between two consequent edges is much greater than such shifts.

\section{Conclusion}
To conclude, we have demonstrated two compact optical realizations of spatial high pass filtering for 1D and 2D variation of wavefronts. Our results exhibit polarization-insensitive performance which permits employment of these filters for unpolarized input profiles. The practical application of these filters for analog optical image processing, specifically edge detection, has been discussed as well. We hope that by using these optical designs as a pre-processing unit, the demand for faster and more efficient computation will be properly satisfied.

\bibliographystyle{ieeetr}
\bibliography{references}

\end{document}